\documentclass[aps,pra,showpacs,twocolumn,superscriptaddress]{revtex4}
\usepackage{amssymb}
\usepackage{amsmath}
\usepackage{graphicx}
\usepackage{epsfig}
\usepackage{subfigure}
\usepackage{xcolor}

\begin{document}

\title{Partially dark optical molecule via phase control}
\author{Z. H. {Wang}}
\affiliation{Center for Quantum Sciences and School of Physics, Northeast Normal
University, Changchun 130024, China}
\affiliation{Center for Advanced Optoelectronic Functional Materials Research, and Key
Laboratory for UV-Emitting Materials and Technology of Ministry of
Education,Northeast Normal University, Changchun 130024, China}
\affiliation{Beijing Computational Science Research Center, Beijing 100094, China}
\author{Yong \surname{Li}}
\email{liyong@csrc.ac.cn}
\affiliation{Beijing Computational Science Research Center, Beijing 100094, China}
\affiliation{Synergetic Innovation Center of Quantum Information and Quantum Physics,
University of Science and Technology of China, Hefei 230026, China}
\affiliation{Synergetic Innovation Center for Quantum Effects and Applications, Hunan
Normal University, Changsha 410081, China}

\begin{abstract}
We study the tunable photonic distribution in an optical molecule consisting
of two linearly coupled single-mode cavities. With the inter-cavity coupling
and two driving fields, the energy levels of the optical-molecule system
form a closed cyclic energy-level diagram, and the phase difference between
the driving fields serves as a sensitive controller on the dynamics of the
system. Due to the quantum interference effect, we can realize a partially
dark optical molecule, where the steady-state mean photon number in one of the cavities achieves zero even under the external driving. And the dark cavity can be changed from one of the cavities to the other by only adjusting the phase difference. Furthermore, we show that when one of the cavities couples
with an atomic ensemble, it will be dark under the same condition as that
without atoms, but the condition for the other cavity to be dark is modified.
\end{abstract}

\maketitle


\section{introduction}

Coherent control of photons is nowadays one of the central topics in quantum
optics and quantum information processing. Usually, the photons are confined
in small volumes~\cite{KJ} of cavities with low dissipation, in order to
promote their controllability and enhance their interaction with matters
such as atoms. Besides, it is convenient to couple the cavities each other
to form controllable quantum network~\cite{HJ} and construct quantum device.
For example, the coupled-cavity array with doped defects can be used to
realize the single-photon transistor or router~\cite{ZL,ZL2,KX,Tian}, and
the system can also be applied to simulate the quantum phase transition in
strong correlated systems~\cite{AD,JK,CD}.

In recent years, the coherent control of photons in a simple system of two
coupled cavities (that is, an optical molecule~\cite{MB,YP}), has invoked a
lot of research interests, including for example, the photon blockade~\cite%
{XW,HS,JL}, state transfer~\cite{CD}, non-equilibrium dynamics~\cite{FN},
coherent polariton~\cite{YCL}, and unidirectional photonic transport~\cite%
{LC,BP}. However these above works appear in the systems of optical
molecules with the assistance of some nonlinear interactions, such as the
Kerr interaction. It naturally inspires us to
investigate the photonic control in the optical-molecule system up to only
the linear terms in the effective interactions.

In this paper, we study the coherent control of mean photon numbers in the
cavities of an optical molecule when it reaches its steady state. In such a
system, one can construct a closed cyclic diagram for the energy levels
(e.g., the lowest three energy ones) with the inter-cavity coupling and two
classical fields which drive the two cavities respectively. And the phase
difference between the two driving fields serves as a sensitive controller
for physical phenomenon. Due to the quantum interference effect, we show
that the partially dark optical molecule can be realized: the steady-state mean photon number of any of the two cavities can be zero (i.e., the cavity is dark since it achieves its steady vacuum state without considering the vacuum fluctuation) even under external driving. Furthermore, the dark cavity can be modified from one cavity to the other by only adjusting the phase difference, which can be easily controlled in experiments.

Furthermore, we consider the case of adding an ensemble of identical atoms
to interact with one of the cavities of the optical molecule. By means of
the bosonization process of the low atomic collective excitations~\cite%
{CollectiveExcitation}, we keep effectively only the linear effects of
atom-cavity interaction and obtain analytically the steady-state values of
the photon numbers in the cavities. Similar to the situation without atomic
ensemble, we can also prepare the partially dark optical molecule with zero
mean photon number in either of the cavities by tuning the phase difference
between the two driving fields. Moreover, compared with the case without
atoms, the condition for the cavity which couples to the atomic ensemble is
not changed but that for the other cavity is significantly modified.

The rest of the paper is organized as follows. In Sec.~\ref{He}, we present
the Hamiltonian and the steady-state values of the photon numbers in the
cavities of the optical-molecule system. The condition to realize partially
dark optical molecule is discussed in Sec.~\ref{ce}. We discuss in Sec.~\ref%
{ha} the realization of partially dark optical molecule in the situation
where one of the cavities is coupled with an ensemble of identical atoms.
A brief conclusion is given in Sec.~\ref{con}.

\section{The Hamiltonian and steady state}

\label{He}

\begin{figure}[tbp]
\begin{centering}
\includegraphics[width=8cm]{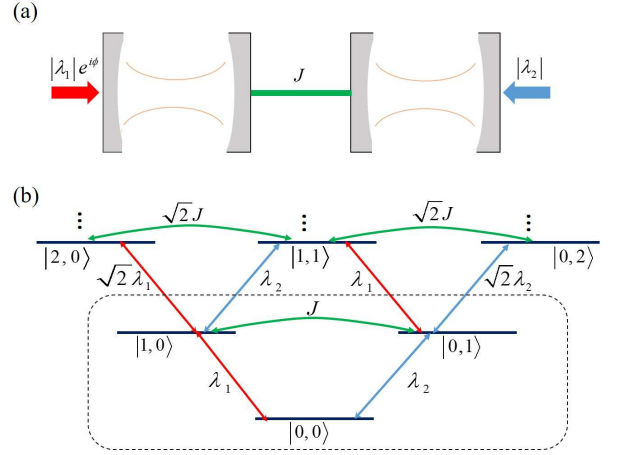}
\end{centering}
\caption{(Color online) (a) Scheme for optical molecule with two driving
fields. (b) Energy-level diagram for an optical molecule with two driving
fields, the energy-levels inside the dashed rectangle frame
forms cyclic transitions.}
\label{scheme}
\end{figure}

The optical-molecule system under consideration is schematically shown in
Fig.~\ref{scheme}(a). The two single-mode cavities couple to each other, and
the Hamiltonian can be written as (here and after $\hbar =1$)
\begin{eqnarray}
H &=&\omega _{1}a_{1}^{\dagger }a_{1}+\omega _{2}a_{2}^{\dagger
}a_{2}+J(a_{1}^{\dagger }a_{2}+a_{2}^{\dagger }a_{1})  \notag \\
&&+(\lambda _{1}a_{1}e^{i\omega _{d}t}+\text{H.c.})+(\lambda
_{2}a_{2}e^{i\omega _{d}t}+\text{H.c.}),
\end{eqnarray}%
where $a_{1}$ ($a_{2}$) is the annihilation operator of cavity mode $1$ ($2$%
) with resonance frequency $\omega _{1(2)}$. $J>0$ is the coupling strength
between the two cavity modes. Furthermore, we introduce a pair of external
classical field with the same frequency $\omega _{d}$ to drive the two
cavities respectively. $\lambda _{1}$ ($\equiv |\lambda _{1}|e^{i\phi }$)
and $\lambda _{2}$ ($\equiv |\lambda _{2}|$) are the driving strength for
the cavities respectively and $\phi $ is the phase difference of the two
driving fields, which can be tuned freely in the regime $-\pi \leqslant \phi
\leqslant \pi $. In the rotating frame with respect to the driving frequency
$\omega _{d}$, the Hamiltonian becomes
\begin{eqnarray}
\mathcal{H} &=&\Delta _{1}a_{1}^{\dagger }a_{1}+\Delta _{2}a_{2}^{\dagger
}a_{2}+J(a_{1}^{\dagger }a_{2}+a_{2}^{\dagger }a_{1})  \notag \\
&&+|\lambda _{1}|(a_{1}e^{i\phi }+a_{1}^{\dagger }e^{-i\phi })+|\lambda
_{2}|(a_{2}+a_{2}^{\dagger }),
\end{eqnarray}%
where $\Delta _{1(2)}=\omega _{1(2)}-\omega _{d}$ is the detuning between
the cavity 1 (2) and the driving field. The dynamics of the system can be
described by the Heisenberg-Langevin equation (neglecting the fluctuations)
\begin{equation}
\dot{\mathbf{A}}=\mathrm{M}\mathbf{{A}+{B},}
\end{equation}%
where $\mathbf{A}=(a_{1},a_{2})^{\text{T}}$, $\mathbf{B}=-i(|\lambda
_{1}|e^{-i\phi },|\lambda _{2}|)^{\text{T}}$, and
\begin{equation}
\mathrm{M}=\left(
\begin{array}{cc}
-(i\Delta _{1}+\frac{\gamma _{1}}{2}) & -iJ \\
-iJ & -(i\Delta _{2}+\frac{\gamma _{2}}{2})%
\end{array}%
\right)
\end{equation}%
with $\gamma _{1(2)}>0$ being the decay rate of cavity mode 1 (2). The
steady-state values of the system are given by
\begin{eqnarray}
\alpha _{1} &=&\langle a_{1}\rangle =-\frac{e^{-i\phi }(R_{1}+iI_{1})}{%
\mathrm{{det}(M)}},  \label{a1} \\
\alpha _{2} &=&\langle a_{2}\rangle =-\frac{(R_{2}+iI_{2})}{\mathrm{{det}(M)}%
},  \label{a2}
\end{eqnarray}%
where
\begin{align}
R_{1}& \equiv |\lambda _{2}|J\cos \phi -|\lambda _{1}|\Delta
_{2},\,I_{1}\equiv \frac{|\lambda _{1}|\gamma _{2}}{2}+|\lambda _{2}|J\sin
\phi ,  \label{c1} \\
R_{2}& \equiv |\lambda _{1}|J\cos \phi -|\lambda _{2}|\Delta
_{1},\,I_{2}\equiv \frac{|\lambda _{2}|\gamma _{1}}{2}-|\lambda _{1}|J\sin
\phi .  \label{c2}
\end{align}

It is obvious that the phase difference $\phi $ plays a significant role in
controlling the photon numbers in the two cavities. The phase dependent
dynamics is ascribed to the quantum interference effect with different
transition paths. As shown in Fig.~\ref{scheme}(b), the diving field of
strength $\lambda _{1}$ induces the transitions $|0,0\rangle \rightarrow
|1,0\rangle \rightarrow |2,0\rangle $ and $|0,1\rangle \rightarrow
|1,1\rangle $, and the driving field of strength $\lambda _{2}$ induces the
transitions $|0,0\rangle \rightarrow |0,1\rangle \rightarrow |0,2\rangle $
and $|1,0\rangle \rightarrow |1,1\rangle $ with $|m,n\rangle $ representing $%
m$ photons in cavity 1 and $n$ photons in cavity 2. Meanwhile, the direct
inter-cavity coupling $J$ induces the transitions $|1,0\rangle \rightarrow
|0,1\rangle $ and $|2,0\rangle \rightarrow |1,1\rangle \rightarrow
|0,2\rangle $. Therefore, a closed cyclic transition forms for any subspace $%
\{|m,n\rangle ,|m,n+1\rangle ,|m+1,n\rangle \}$ {(for example, see the states in the dashed rectangle frame for the case of $m=n=0$)} and the relative total phase of the loop $\phi$ will essentially influence the steady state of the system.
However, when any of the driving fields is shut down, the closed transition
disappears and the phase difference $\phi $ will take no effect in
controlling the average photon numbers in the cavities. This fact can also
be observed in Eqs.~(\ref{a1}-\ref{c2}), which show that both $|\alpha
_{1}|^{2}$ and $|\alpha _{2}|^{2}$ are independent of the phase difference $%
\phi $ when $|\lambda _{1}|=0$ or $|\lambda _{2}|=0$. Actually, the similar
closed cyclic energy-level diagram can also be found in many other systems,
such as superconducting artificial atom~\cite{YL,WJ}, chiral molecule \cite%
{ChiralMolecule,ChiralMolecule2}, cavity-QED system~\cite{JT}, and cavity
optomechanical system~\cite{WJ1,XW3}, in which the phase control to quantum
phenomenon has attracted much attention.

\begin{figure}[tbp]
\centering
\includegraphics[width=8cm]{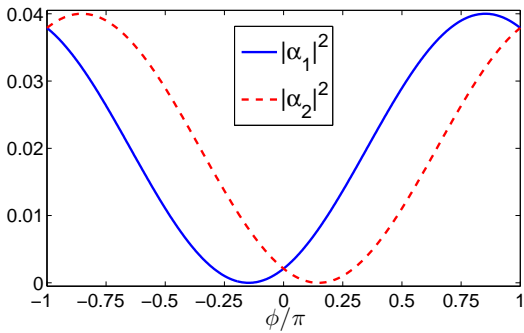}
\caption{(Color online) The average photons number in cavity 1 ($|\protect%
\alpha _{1}|^{2}$) and cavity 2 ($|\protect\alpha _{2}|^{2}$) as functions
as the relative phase of the two driving fields $\protect\phi $. The
parameters are set as $\protect\gamma _{1}=\protect\gamma _{2}=\protect%
\gamma ,\Delta _{1}=\Delta _{2}=\protect\gamma ,|\protect\lambda _{1}|=|%
\protect\lambda _{2}|=0.1\protect\gamma ,J=\protect\sqrt{5}\protect\gamma /2$%
. Under these parameters, the conditions in Eqs.~(\protect\ref{co1},\protect
\ref{co2}) are both satisfied.}
\label{n1n2}
\end{figure}

In Fig.~\ref{n1n2}, we plot the average photon numbers $|\alpha _{1}|^{2}$
and $|\alpha _{2}|^{2}$ as functions of the phase difference $\phi $. It is
interesting that, even in the presence of the driving fields, we can still
achieve the regime in which $|\alpha _{1}|^{2}=0$ or $|\alpha _{2}|^{2}=0$.
It implies that we can realize the partially dark optical molecule, i.e.,
one of the cavities will stabilize in its vacuum state, and thus will be
named as dark cavity in the following. The parameter condition to realize
such a partially dark optical molecule will be discussed in the next section.

\section{Realization of partially dark optical molecule}
\label{ce}

In the last section, we have shown that by adjusting the phase difference
between the two driving fields, one of the cavities can be tuned to be in
its steady vacuum state. In this section, we will give the conditions for
realizing such a partially dark optical molecule.

First, we seek the conditions for the average photon number of the cavity
mode 1 being zero ($\alpha _{1}=0$). As shown in Eqs.~(\ref{a1},\ref{c1}),
this requires $R_{1}=I_{1}=0$, which yields
\begin{equation}
\cos \phi =\frac{|\lambda _{1}|\Delta _{2}}{|\lambda _{2}|J},\,\sin \phi =-%
\frac{|\lambda _{1}|\gamma _{2}}{2|\lambda _{2}|J}.  \label{phi1}
\end{equation}%
The above equation implies
\begin{equation}
|\lambda _{2}^{2}|J^{2}=|\lambda _{1}^{2}|(\Delta _{2}^{2}+\frac{\gamma
_{2}^{2}}{4}).  \label{co1}
\end{equation}

Similarly, the conditions for $\alpha _{2}=0$ can be expressed as
\begin{equation}
\cos \phi =\frac{|\lambda _{2}|\Delta _{1}}{|\lambda _{1}|J},\,\sin \phi =%
\frac{|\lambda _{2}|\gamma _{1}}{2|\lambda _{1}|J},  \label{phi2}
\end{equation}%
which implies
\begin{equation}
|\lambda _{1}^{2}|J^{2}=|\lambda _{2}^{2}|(\Delta _{1}^{2}+\frac{\gamma
_{1}^{2}}{4}).  \label{co2}
\end{equation}

From the above conditions, we note that the steady photon number in one of
the two cavities can be zero (in other words, to realize the partially dark
optical molecule) by tuning the parameters of the system. Especially, as
shown in Fig.~\ref{n1n2}, when the parameters are set such that $|\lambda
_{1}|=|\lambda _{2}|=\lambda >0$, $\Delta _{1}=\Delta _{2}=\Delta $, $\gamma
_{1}=\gamma _{2}=\gamma >0$, and $J=\sqrt{\Delta ^{2}+\gamma ^{2}/4}$, we
can transfer the dark cavity from cavity 1 to cavity 2 only by adjusting the
phase difference adiabatically while keeping the other parameters unchanged, which is
available in realistic physical systems.

Here, we emphasize that the two cavities of the optical molecule can not
stabilize in their vacuum states simultaneously (i.e., $\alpha _{1}=\alpha
_{2}=0$). This can be observed from Eqs.~(\ref{phi1},\ref{phi2}), which
imply $\sin \phi <0$ when $\alpha _{1}=0$ and $\sin \phi >0$ when $\alpha
_{2}=0$. Thus one can not achieve $\alpha _{1}=\alpha _{2}=0$ in this case.
Intuitively speaking, this contradiction may be expected to disappear in the
system with balanced loss and gain in which $\gamma _{1}=-\gamma _{2}>0$~%
\cite{JL,BP,LC}. However, further calculation shows $\mathrm{det}(\mathrm{M}%
)=0$ in such a situation if Eqs. (\ref{phi1}-\ref{co2}) are satisfied. That
means there does not exist the steady state in the system. In what follows,
we will just consider the damping case with $\gamma _{1}=\gamma _{2}=\gamma
>0$.

\section{Partially dark optical molecule with atomic ensemble}

\label{ha}

\begin{figure}[tbp]
\centering
\includegraphics[height=5cm,width=8cm]{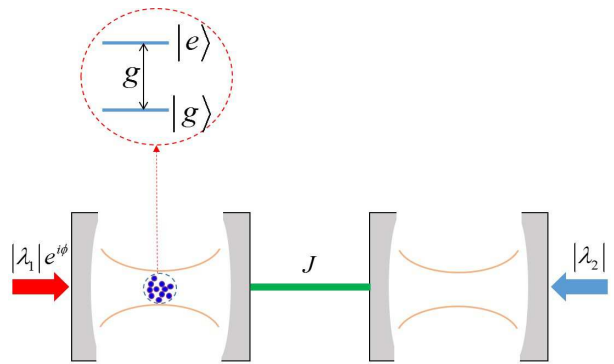}
\caption{(Color online) Scheme for an ensemble of identical two-level atoms
trapping in the optical molecule.}
\label{atomOM}
\end{figure}

Based on the above discussions about the realization of partially dark
optical molecule in the last section, here we continue to study the effects
of atom-cavity interaction in the system. To this end, we now consider an
ensemble of $N$ identical two-level atoms trapped in cavity 1 as shown in
Fig.~\ref{atomOM}.

In the rotating frame with respect to the frequency $\omega _{d}$ of the
driving field, the Hamiltonian under consideration reads
\begin{eqnarray}
\mathcal{H}^{\prime } &=&\Delta _{1}a_{1}^{\dagger }a_{1}+\Delta
_{2}a_{2}^{\dagger }a_{2}+\frac{\Delta _{b}}{2}\sum_{i}^{N}\sigma _{z}^{(i)}
\notag \\
&&+g\sum_{i}^{N}(a_{1}^{\dagger }\sigma _{-}^{(i)}+a_{1}\sigma
_{+}^{(i)})+J(a_{1}^{\dagger }a_{2}+a_{2}^{\dagger }a_{1})  \notag \\
&&+|\lambda _{1}|(a_{1}^{\dagger }e^{-i\phi }+a_{1}e^{i\phi })+|\lambda
_{2}|(a_{2}^{\dagger }+a_{2}),  \label{Hatom}
\end{eqnarray}%
where $\Delta _{b}=\omega _{0}-\omega _{d}$ with $\omega _{0}$ the energy
level spacing between the ground state $|g\rangle $ and excited state $%
|e\rangle $ of the two-level atoms, $g$ is the coupling strength between
single atom and cavity mode 1. $\sigma _{z}^{(i)}=|e\rangle _{ii}\langle
e|-|g\rangle _{ii}\langle g|$ and $\sigma _{+}^{(i)}=[\sigma
_{-}^{(i)}]^{\dagger }=|e\rangle _{ii}\langle g|$ are the Pauli operators
for the $i$-th atom.

To simplify the above Hamiltonian (\ref{Hatom}), we introduce the collective
operators for the atomic ensemble,
\begin{equation}
b=\frac{1}{\sqrt{N}}\sum_{i}^{N}\sigma _{-}^{(i)},\,b^{\dagger }=\frac{1}{%
\sqrt{N}}\sum_{i}^{N}\sigma _{+}^{(i)}.
\end{equation}

In the low-excitation limit with large $N$, the above operators satisfy the
standard commutation relation of the bosonic operators \cite%
{CollectiveExcitation}
\begin{equation}
\lbrack b,b^{\dagger }]\approx 1.
\end{equation}%
Furthermore, we also have
\begin{equation}
\sum_{i}\sigma _{z}^{(i)}=2b^{\dagger }b-N.
\end{equation}%
In terms of the collective operators $b$ and $b^{\dagger }$, the
Hamiltonian~(\ref{Hatom}) can be written as
\begin{eqnarray}
\mathcal{H}^{\prime } &=&\Delta _{1}a_{1}^{\dagger }a_{1}+\Delta
_{2}a_{2}^{\dagger }a_{2}+\Delta _{b}b^{\dagger }b  \notag \\
&+&(Ja_{1}^{\dagger }a_{2}+\eta a_{1}^{\dagger }b+|\lambda
_{1}|a_{1}e^{i\phi }+|\lambda _{2}|a_{2}+\text{H.c.}), \ \ \  \label{Hatom2}
\end{eqnarray}%
where $\eta =g\sqrt{N}$ is the collective coupling strength between the
atomic ensemble and cavity mode 1. Here, we have neglected the constant term
$-N\Delta _{b}/2$.

Based on the above Hamiltonian (\ref{Hatom2}), the Heisenberg-Langevin
equation of the system can be written as
\begin{equation}
\dot{\mathbf{A}^{\prime }}=\mathrm{\mathrm{M}^{\prime }}\mathbf{A}^{\prime }+%
\mathbf{B}^{\prime },
\end{equation}%
where $\mathbf{A}^{\prime }=(a_{1},a_{2},b)^{\text{T}}$, $\mathbf{B}^{\prime
}=-i(|\lambda _{1}|e^{-i\phi },|\lambda _{2}|,0)^{\text{T}}$, and
\begin{equation}
\mathrm{\mathrm{M}^{\prime }}=\left(
\begin{array}{ccc}
-(i\Delta _{1}+\frac{\gamma _{1}}{2}) & -iJ & -i\eta \\
-iJ & -(i\Delta _{2}+\frac{\gamma _{2}}{2}) & 0 \\
-i\eta & 0 & -(i\Delta _{b}+\frac{\gamma _{b}}{2})%
\end{array}%
\right)
\end{equation}%
with $\gamma _{b}$ the decay rate for the atomic ensemble.

The steady state values of the operators are obtained as
\begin{subequations}
\label{ss-atom}
\begin{eqnarray}
\alpha _{1}^{\prime } &=&\langle a_{1}\rangle ^{\prime }=\frac{e^{-i\phi
}(i\Delta _{b}+\frac{\gamma _{b}}{2})(R_{1}+iI_{1})}{-\mathrm{{det}(\mathrm{M%
}^{\prime })}},  \label{ah} \\
\alpha _{2}^{\prime } &=&\langle a_{2}\rangle ^{\prime }=\frac{(i\Delta _{b}+%
\frac{\gamma _{b}}{2})(R_{2}+iI_{2})+i|\lambda _{2}|\eta ^{2}}{-%
\mathrm{{det}(\mathrm{M}^{\prime })}}, \label{bh}\\
\beta ^{\prime } &=&\langle b\rangle ^{\prime }=\frac{i\eta }{i\Delta _{b}+%
\frac{\gamma _{b}}{2}}\alpha _{1}^{\prime }.  \label{beta}
\end{eqnarray}%
\end{subequations}
Here the superscript ``${\prime }$'' denotes for the case with atoms.

From the results given by Eq.~(\ref{ss-atom}), we observe the following
three points. First, the condition for vacuum steady state in cavity 1 is
not changed whenever the atomic ensemble couples or does not couple with the
cavity mode by comparing Eq. (\ref{ah}) with Eq. (\ref{a1}). Second, as
shown in Eqs.~(\ref{ah},\ref{beta}), when the cavity 1 is in the vacuum state ($%
\alpha _{1}^{\prime }=0$), it will not excite the atomic ensemble coupling
with it, that is $\beta ^{\prime }=0$. Third, the condition for the cavity mode
2 achieving its vacuum steady state ($\alpha _{2}^{\prime }=0$) is modified
due to the coupling between the atomic ensemble and the cavity mode 1 [As shown in Eq.~(\ref{co2}) and Eq.~(\ref{bh})]. A
direct calculation shows that, when $\alpha _{2}^{\prime }=0$, we need

\begin{subequations}
\label{phil2atoms}
\begin{eqnarray}
\cos \phi  &=&\frac{\Delta _{1}|\lambda _{2}|(\Delta _{b}^{2}+\gamma
_{b}^{2}/4)-\Delta _{b}|\lambda _{2}|\eta ^{2}}{J|\lambda _{1}|(\Delta
_{b}^{2}+\gamma _{b}^{2}/4)}, \\
\sin \phi  &=&\frac{|\lambda _{2}|\gamma _{1}(\Delta _{b}^{2}+\gamma
_{b}^{2}/4)+\gamma _{b}|\lambda _{2}|\eta ^{2}}{2J|\lambda _{1}|(\Delta
_{b}^{2}+\gamma _{b}^{2}/4)},
\end{eqnarray}%
which imply
\end{subequations}
\begin{equation}
\left[ \Delta _{1}-\frac{\Delta _{b}\eta ^{2}}{(\Delta _{b}^{2}+\gamma
_{b}^{2}/4)}\right] ^{2}+\left[ \frac{\gamma _{1}}{2}+\frac{\gamma _{b}\eta
^{2}}{2(\Delta _{b}^{2}+\gamma _{b}^{2}/4)}\right] ^{2}
=\frac{J^{2}|\lambda _{1}|^{2}}{|\lambda _{2}|^{2}}.
\label{cat}
\end{equation}

\begin{figure}[tbp]
\centering
\includegraphics[width=8cm]{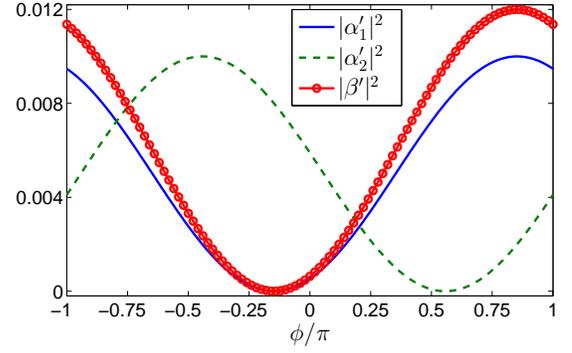}
\caption{(Color online) The average photons number in cavity 1 ($|\protect%
\alpha_1^{\prime}|^2$), cavity 2 ($|\protect\alpha_2^{\prime}|^2$) and the excitation of the
atomic ensemble ($|\protect\beta^{\prime}|^2$) as functions of the phase difference
of the two driving fields $\protect\phi$. The parameters are set as $\protect%
\gamma_1=\protect\gamma_2=\protect\gamma_3=\protect\gamma,\Delta_1=\Delta_2=%
\Delta_3=\protect\gamma,\protect\lambda_1=\protect\lambda_2=0.1\protect\gamma%
,g=\protect\sqrt{5}\protect\gamma/2,\protect\eta=\protect\sqrt{6}\protect%
\gamma/2$. Under these parameters, the conditions in Eqs.~(\protect\ref{co1},%
\protect\ref{cat}) are both satisfied.}
\label{n1n2atom}
\end{figure}

Similar to the situation without atoms, in the situation with atoms we can
also transfer the dark cavity from one cavity to the other by only tuning
the phase difference between the two driving fields adiabatically while keeping the other
parameters unchanged. That is, when the conditions in Eq.~(\ref{co1}) and
Eq.~(\ref{cat}) are fulfilled simultaneously, one can modify the phase
difference $\phi $ to satisfy Eq. (\ref{phi1}) in order to make the cavity 1
dark ($\alpha _{1}^{\prime }=0$) or satisfy Eqs. (\ref{phil2atoms}) to make
the cavity 2 dark ($\alpha _{2}^{\prime }=0$). When appropriately choosing
the parameters such that $|\lambda _{1}|=|\lambda _{2}|=\lambda $, $\Delta
_{1}=\Delta _{2}=\Delta _{b}=\Delta $, and $\gamma _{1}=\gamma _{2}=\gamma
_{b}=\gamma $, the above two conditions in Eqs. (\ref{co1},\ref{cat}) are
simply expressed as
\begin{equation}
J=\sqrt{\Delta ^{2}+\gamma ^{2}/4},\,\eta =\sqrt{2(\Delta ^{2}-\gamma ^{2}/4)}.
\end{equation}

Under these conditions, we plot the average photon numbers in the two
cavities and the excitation number of the atomic ensemble as functions of
the phase difference $\phi $ between the two driving fields in Fig.~\ref%
{n1n2atom}, with the parameters setting as $|\lambda _{1}|=|\lambda
_{2}|=\lambda =0.1\gamma $, $\Delta _{1}=\Delta _{2}=\Delta _{b}=\Delta
=\gamma $, $g=\sqrt{5}\gamma /2$, and $\eta =\sqrt{6}\gamma /2$. As shown in
this figure, when $\phi =-0.14\pi $ [$=\arcsin (-1/\sqrt{5})$], the cavity
mode 1 achieves the vacuum steady state, and the atomic ensemble will not be
excited ($|\alpha _{1}^{\prime }|^{2}=|\beta ^{\prime }|^{2}=0$). On the
other hand, when $\phi $ is tuned to $\phi =0.557\pi $ [$=\arccos (-0.08%
\sqrt{5})$], the cavity mode 2 can be stabilized in its vacuum state ($%
|\alpha _{2}^{\prime }|^{2}=0$), while the cavity mode 1 and atomic ensemble
are excited.


\section{Conclusion}

\label{con}

In summary, we have shown the scheme to realize the partially dark optical
molecule via only tuning the phase difference between the two driving
fields. The fact that the avearge photon numbers of the cavities in the
optical molecule are significently dependent on the phase difference results
from the quantum interference effect happening in the colsed energy level
diagram. We analytically give the conditions to realize the partially dark
molecule (that is one of the cavity modes achieves its vacuum steady state).
Moreover, when an additional ensemble of two-level atoms is coupled with one
of the cavities (e.g. cavity 1), analytical calculation showed that the
optical molecule can be still partially dark.\textit{\ }In both of the
situations when the atomic ensemble is present or absent, we find that the
dark cavity can be transferred from one cavity to the other only by
adjusting the phase difference while keeping other parameters remained.
Compared with the case without atoms, the condition for cavity 1 being dark
is unchanged and that for cavity 2 being dark is modified in the case with
atoms. It is interesting that when the cavity 1 is dark, the atomic ensemble
inside cavity 1 will also be dark with 0 excitation number. That means the
cavity as well as the atoms inside it can be \textquotedblleft
shielded\textquotedblright\ in the optical molecule even at the present of
the optical drivings. Our scheme for phase control in optical
molecules might provide a platform for applications on quantum information
process based on photonic devices.

\begin{acknowledgments}
This work is supported by the National Basic Research Program of China (under Grants No. 2014CB921403 and No. 2016YFA0301200), NSFC (under Grants No. 11404021 and No. U1530401), and the Fundamental Research Funds for the Central Universities (under Grant No. 2412016KJ015).

\end{acknowledgments}

\end{document}